\documentclass[prb,showpacs,showkeys,preprintnumbers,superscriptaddress,a4paper,twocolumn]{revtex4}
\usepackage[T1]{fontenc}
\usepackage[utf8]{inputenc}
\usepackage{graphicx}
\usepackage{lmodern}
\usepackage[english]{babel}
\usepackage{amsmath,amssymb}
\usepackage{amstext}
\usepackage[colorlinks=true,linkcolor=red]{hyperref}
\usepackage{color}														
\usepackage{graphicx}

\newcommand{\vs}{\vec{s}}

\newcommand{\etx}{\tilde{\mathbf{e}}_x}
\newcommand{\etz}{\tilde{\mathbf{e}}_z}
\newcommand{\ety}{\tilde{\mathbf{e}}_y}

\begin{document}
\title{Incommensurate, helical spin ground states on the Hollandite lattice}
\author{S. Mandal}\email{smandal1@ictp.it}\affiliation{The Abdus Salam ICTP, Strada Costiera 11, I-34151 Trieste, Italy}
\author{A. Andreanov}\email{alexei@pks.mpg.de}\affiliation{Max-Planck-Institut f\"ur Physik komplexer Systeme, N\"othnitzer Str. 38, 01187 Dresden, Germany}
\author{Y. Crespo}\email{ycrespo@ictp.it}\affiliation{The Abdus Salam ICTP, Strada Costiera 11, I-34151 Trieste, Italy}
\author{N. Seriani}\email{nseriani@ictp.it}\affiliation{The Abdus Salam ICTP, Strada Costiera 11, I-34151 Trieste, Italy}
\begin{abstract}
	We present a model of classical Heisenberg spins on a Hollandite lattice, which has been developed to describe the magnetic
	properties of $\alpha$-MnO$_2$ and similar compounds. The model has nearest neighbor interacting spins, however the strength
	and the sign of spin-spin interactions is anisotropic and depends on the nature of the bonds. Our analysis shows that the Hollandite lattice supports four different incommensurate and helical magnetic ground
	states depending on the relative strengths and signs of spin-spin interactions. We show that the incommensurate helical ground 
	states appear due to the geometrical frustration present in the model. We demonstrate that each of the four helical incommensurate
	magnetic phases are continuously connected to four different collinear antiferromagnetic ground states 
	as the strength of spin-spin interaction along some bonds is increased. The present results give support to the presence of helical states that have been previously suggested experimentally 
for Hollandite  compounds. We provide an in-depth analysis of the magnetic form factors for each helical phase and describe how it could be used to identify each of these phases in neutron diffraction experiments.
\end{abstract}

\date{\today}

\maketitle

\section{Introduction}
\label{sec:intro}

Magnetic compounds with a low magnetic anisotropy are interesting for the large variety of magnetic phases they display.  In general, the starting point to understand these magnetic phases are interacting Heisenberg spins defined on lattices that mimic the lattice structure of the actual compounds. While examining quantum Heisenberg spins remains a challenging task, ~\cite{balents2010frustration,powell2011review} in many cases, understanding the phase diagram for interacting classical Heisenberg spins is also a complicated problem.~\cite{diep2004spin} Actually, it has been proven that geometrical frustration combined with complex lattice structure can make such analysis highly non-trivial. In many cases various exotic magnetic ground states were identified, for example, for pyrochlore  lattices~\cite{reimers1991pyrochlores,moessner1998pyrochlore,moessner1998frustrated}, diamond lattice,~\cite{fritsch2004diamond,
krimmel2006diamond, bergman2007diamondafm} {\it etc.}. One of the key ingredients of geometrical frustration is the presence of triangles as structural units in a  lattice.~\cite{anderson1973resonating,fazekas1974on} For this reason the study of antiferromagnetic Heisenberg spin models on various triangular lattices, including one dimensional ladder, $2$D and $3$D triangular lattices, and quasi-$2$D triangular ladders, has drawn considerable attention in the scientific community.~\cite{rajiv1992tr-kagome,weihong1999trafm,bernu1992finitelat,chen2013trlatt-magfield}

\begin{figure}[!tbp]
\begin{center}
 	 	\includegraphics[width=0.9\columnwidth]{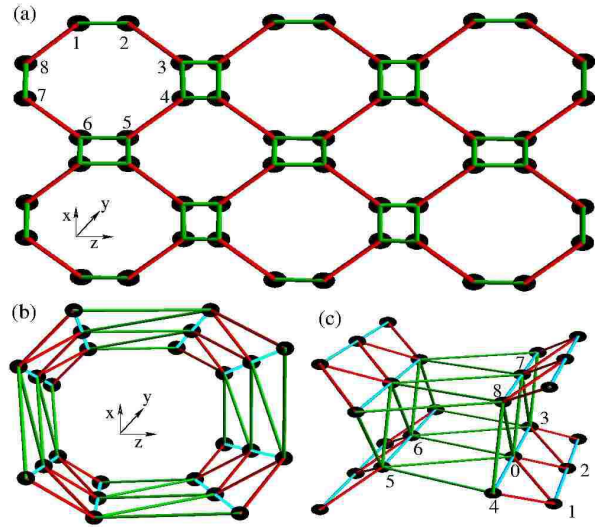}
	\caption{(Color online) Different views of the Hollandite lattice. In this picture, the black points denote	the position of the transition metal atom. (a) An $x-z$ plane projection including the conveniently used $8$-atom unit cell enumerated in the clockwise direction. (b) Panoramic view of $8$-atom channel extended in	$y$-directions. (c)  Panoramic view  of $4$-atom channel, connecting $4$ ladders. We also show the $8$ nearest neighbors of site $0$ that are labeled by numbers $1..8$.}
	\label{fig:MnO2-lattice}
\end{center}
\end{figure}

One example of such lattice with geometrical frustration is the Hollandite lattice,~\cite{deguzman1994synthesis,suib1994magnetic} which
is the crystal structure of some transition metal oxides as in the case of $\alpha$-MnO$_2$.~\cite{hasegawa2009discovery, deguzman1994synthesis, 
ishiwata2006structure} This manganese oxide has recently attracted considerable attention from both experimental and theoretical
condensed matter  research communities~\cite{yamamoto1974single,luo2010tuning,luo2009spin,shen2005a,sato1999charge,sato1997tunnel,cockayne2012first,crespo2013competing,
crespo2013electronic,tompsett2013electrochemistry} because of its large number of applications as a catalyst for oxygen reduction 
reaction~\cite{Appl.ORR}, microbial fuel cells~\cite{Appl.fuel.cells}, electrode materials for Li-ion batteries~\cite{thackeray1997manganese}, 
lithium-air batteries~\cite{bruce2012lio2,girishkumar2010lithium,song2011nanostructured, tompsett2013electrochemistry,thackeray2013} and in 
supercapacitors~\cite{li2010mesoporous}, \textit{etc.}. Hollandite-type compounds display pores with 
a diameter of $\sim$4.6 \AA, (see Fig.\ref{fig:MnO2-lattice}(a)). These large channels can accommodate cations such 
as K$^+$, Na$^+$, and Ba$^{2+}$, leading to non-stoichiometric compounds such as
K$_{1.5}$(H$_3$O)$_x$Mn$_8$O$_{16}$~\cite{deguzman1994synthesis,suib1994magnetic,sato1999charge} and Ba$_{1.2}$Mn$_8$O$_{16}$.
~\cite{ishiwata2006structure} The presence of these impurities makes it hard to determine their composition, in particular the water 
content.~\cite{gao2008microstructures,sato1999charge,sato1997tunnel,strobel1984thermal} This uncertainty in the composition introduces a
difficulty for studying the electronic and magnetic properties of these compounds, as they vary depending on the type of dopant and its concentration. The magnetic frustration due to the structural triangles (see Fig.\ref{fig:MnO2-lattice}(b)), and the presence of impurities in the channels are at the origin of the rich variety of magnetic ordered phases that have been identified experimentally in these materials. The 
best studied case is that of potassium impurities. Strobel \textit{et al.} in Ref.~\onlinecite{strobel1984thermal} reported an antiferromagnetic (AFM) transition for K$_{0.16}$MnO$_2$ at T$_N$=$18$ K, also an AFM ground state was reported for K$_{<0.7}$MnO$_2$ synthesized with a hydrothermal technique.~\cite{yamamoto1974single} However, the exact nature of this AFM phase is still unknown. Increasing the amount of K$^+$ ions in the channels of K$_x$MnO$_2$ ($0.087<x\leq 0.125$) a spin-glass behavior was observed at low temperatures.~\cite{luo2010tuning,luo2009spin,shen2005a} Moreover, impurities can further complicate the phase diagram: the K$_{1.5}$(H$_3$O)$_x$Mn$_8$O$_{16}$ exhibits a ferromagnetic (FM) transition at $52$ K and this FM state persists down to $20$ K. When the temperature is further lowered, the ferromagnetic state disappears and possibly a helical magnetic state is established, as suggested by the spatial anisotropic behavior of the susceptibility.~\cite{sato1999charge} A helical magnetic structure was also suggested 
for K$_{0.15}$MnO$_2$ at low temperatures.~\cite{sato1997tunnel}

Recently some theoretical studies have been carried out~\cite{crespo2013competing,crespo2013electronic} in order to understand the magnetic phase diagrams of the Hollandite type magnetic compounds. In particular in Ref.~\onlinecite{crespo2013competing} Crespo and co-workers used a simple Ising model to qualitatively reproduce both the antiferromagnetic ground state~\cite{yamamoto1974single} in the absence of disorder and the spin-glass transition under doping.~\cite{luo2010tuning,luo2009spin,shen2005a} Nevertheless this study was only a zeroth order approximation to this type of systems as manganese compounds are known to display low magnetic anisotropy. In fact experimental data for several manganese compounds ~\cite{Moussa1996,Chaboussant2004,Fabreges2011} have shown that these compounds are well described by the Heisenberg model. Consequently, the model of Ref.~\onlinecite{crespo2013competing} cannot reproduce realistic low temperature magnetic configurations of manganites, like non-collinear AFM or helical 
magnetic ground states, that have been suggested in some compounds.~\cite{sato1997tunnel,sato1999charge}

In this paper we extend the previous work~\cite{crespo2013competing} and study the classical Heisenberg model on the Hollandite lattice in order to investigate the ground state magnetic phase diagram as a function of the model parameters. By exploring all  possible   magnetic interactions between the spins we probe different uniform compositions of the Hollandite structure and therefore give an explanation to some of the magnetic ground states that have been experimentally suggested. We have found that the Hollandite lattice supports many different ground states inside the phase space of model parameters: from various collinear AFM spin configurations, as in the Ising case~\cite{crespo2013competing} for large values of the AFM couplings, to incommensurate helical spin configurations when all the interactions have comparable strengths. To find the ground states for a given set of couplings two different approaches have been used: (i) the  well-known interaction matrix analysis~\cite{luttinger1946ltz,
litvin1974ltz} and (ii) numerical simulations~\cite{sklan2013nonplanar} employing  local field quench method.~\cite{walker1980computer} Both methods agreed perfectly
and yielded identical ground state energy. 

The paper is organized as follows: we introduce the model in Sec.\ref{sec:model} and the ground states for different choices of the interaction parameter  are discussed in Sec.\ref{sec:clean}. In Sec.\ref{sec:swt-et}, the magnetic structure factor is  analysed in detail for each region of the phase diagram. Finally, the effect of an uniform external magnetic field is discussed in Sec.\ref{magfield-effect}. We conclude our work by summarizing our results.

\section{Model}
\label{sec:model}

The Hollandite structure is shown schematically in Fig.\ref{fig:MnO2-lattice}. The lattice can be described as a collection of connected planes where each plane is extended parallel to the $x-z$ plane and is interconnected to the other planes in $y$-direction. The conventional unit cell is composed by $8$ sites that have been numerated in clockwise direction in the upper left corner of Fig.\ref{fig:MnO2-lattice}(a) where the $x-z$ projection of the lattice is shown. Repetition of the unit cell in the $y$-direction creates a channel-like structure that together with its local triangular form are the main features of the lattice. The connectivity between nearest unit cells extended in $y$-directions is shown in Fig.\ref{fig:MnO2-lattice}(b). A given site of the Hollandite lattice has $8$ nearest neighbors, that are shown in Fig.\ref{fig:MnO2-lattice}(c). We remark that in real materials, like the $\alpha$-MnO$_2$ compounds, these $8$ neighbors are at different distances, the length of the bonds 
being $\sim$2.86 \AA, 2.91 \AA ~and 3.44 \AA, respectively. Nevertheless these sites have to be considered as nearest neighbors since removing the longer links disrupts the lattice connectivity.

\begin{figure}[!t]
\begin{center}
 	\includegraphics[width=0.9\columnwidth]{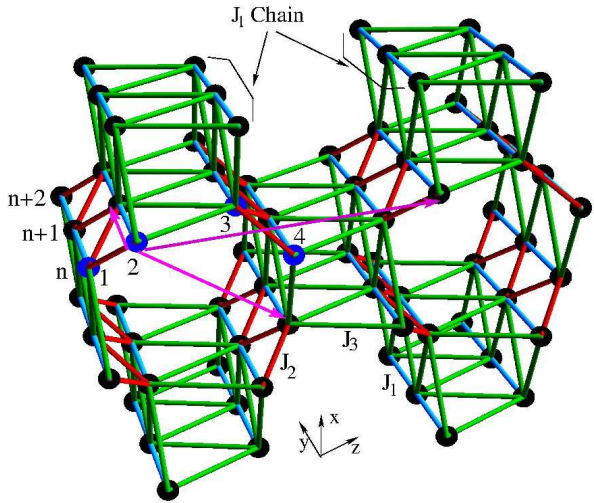}
	\caption{(Color online) A view of Hollandite lattice structure which shows  the distribution of ($J_1,J_2,J_3$) bonds for different kind of nearest neighbours.  The blue lines extending in $y$-directions denote $J_1$ bonds. The red lines denote the $J_2$ bonds and the green links are the $J_3$ bonds. This figure also shows the definition of a $J_1$ chain and how the lattice sites $n,n+1,n+2$ inside this chains are labeled. Finally it is shown that this lattice could also be described with 4-site unit cell, where the sites selected in this work are represented by blue points and the three lattice vectors $\vec{A}_1,\vec{A}_2,\vec{A}_3$ for this representation are also given as pink arrows. The details of the lattice parameters are given in Appendix\ref{sec:int-mat}.}
	\label{fig:MnO2-interactions}
\end{center}
\end{figure}

The magnetism in $\alpha$-MnO$_2$ materials is due to the interaction of the magnetic moments localized on manganese ions, which interact with each other through oxygen-mediated super-exchange.~\cite{sato1999charge,umek2009synthesis,kanamori1959superexchange,goodenough1955theory,goodenough1958an} We consider here the simplest possible model compatible with the lattice structure and the low magnetic anisotropy of manganese ions on manganese oxides: we place classical Heisenberg spins on the sites of the Hollandite lattice. We restrict the interaction to nearest neighbors. In fact recent results, based on density functional theory show that the contribution of the second nearest-neighbors to the total energy of the system is negligible ~\cite{crespo2013electronic}. 
We therefore consider a model similar to the one of Ref.~\onlinecite{crespo2013competing} where three couplings strengths - $J_1, J_2$, $J_3$ - were used for the spin-spin interactions, one for each of the three types of nearest neighbors. Assignment of $J_1,~J_2$ and $J_3$ is shown in Fig.\ref{fig:MnO2-interactions}. Therefore the Heisenberg Hamiltonian reads
\begin{gather}
	\small
	\label{eqn:H}
	\mathcal{H} = \sum_{k=1}^{3}\sum\limits_{<ij>_k} J_{k}^{ij}\vs_i\cdot\vs_j,
\end{gather}
where $<ij>_k$, $k$ denotes three different groups of nearest neighbors of spin `$i$'. We used three colors - $J_1$ blue, $J_2$ red and $J_3$ green - in all figures to represent each type of nearest neighbors. Positive values of the couplings correspond to AFM interactions, the negative values refer FM interactions. We remark that such choice of the couplings makes the interactions $J$-anisotropic in general.
 
Despite the seeming simplicity of the Hamiltonian~\eqref{eqn:H}, as we find below, it admits many different ordered magnetic ground states depending on signs and relative strengths of the couplings $J_k$. This richness reflects the complicated geometry of the Hollandite lattice and the frustration present for some values of the $J_k$ couplings. Since in Hollandite-type structures the angles of Mn-O-Mn bridges take values between $80$ and $130$ degrees, simple symmetry rules that determine the sign of the magnetic couplings, such as the Goodenough-Kanamori  rules,~\cite{goodenough1955theory,goodenough1958an,kanamori1959superexchange} can not be applied. We failed to find any experimental insights for the selection of particular values of the couplings $J_k$. In general, we expect
a different set of couplings for each material. Therefore, in this study, we describe the possible magnetic ground states of the Hamiltonian~\eqref{eqn:H} for all possible values of the coupling 
strengths $J_k$. This also provides us useful insights regarding the details of the couplings (sign and the value) that a particular compound may have.

\section{Phase diagram}
\label{sec:clean}

In this section we present the ground states of the Hamiltonian~\eqref{eqn:H}, assuming that all the couplings $J_k$ are constant and do not vary in space 
and we chart out the phase diagram for all possible combinations of signs and relative strengths of the couplings. The phase diagram depends
crucially on the sign of the coupling $J_1$. Since the Hollandite lattice is composed of triangular ladders (see Fig.\ref{fig:MnO2-interactions}),
the sign of $J_1$ determines whether the ladders are frustrated (the case of $J_1 > 0$) or not (the case of $J_1 < 0$). This observation allows
us to study the two cases separately and plot the phase diagram as a function of the ratios $J_2/J_1$ and $J_3/J_1$. The case $J_1<0$ renders the 
spin-spin interactions unfrustrated, so that only simple FM or AFM ground states are present, depending on the sign of $J_2$ and $J_3$. Such phase
diagram is similar to that of the Ising spins.~\cite{crespo2013competing}

\begin{figure}[!t]
\includegraphics[width=0.9\columnwidth]{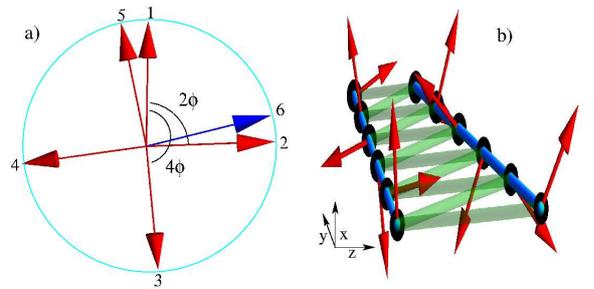}
	\caption{(Color online) Non-commensurate helical spin configuration with a rotation angle of $2\phi$ in the $y$ direction. (a) Projection of the first 6 spins on the $x-z$ plane where the non-commensurate nature of the ground state have been shown for an angle of $\phi=0.76 ~{\rm radian}$. (b) Propagation of the helical spin configuration in the $y$-direction along the two neighboring $J_1$-chains.} 
	\label{fig:helical}
\end{figure}

\begin{figure}[!t]
 	\includegraphics[width=0.99\columnwidth]{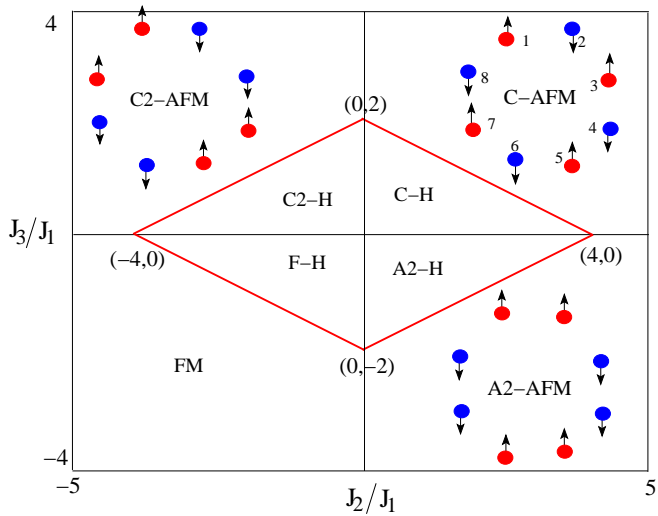}
	\caption{(Color online) Phase diagram obtained for the Hamiltonian given by~\eqref{eqn:H} for $J_1>0$. The region inside the red line correspond to the four co-planar helical spin ground states configuration found in this work, C-H, C2-H, A2-H and F-H (see text and Fig.\ref{fig:xzhelical}). The corner region of each quarter (large values of the couplings) refer to the ordered phases C-AFM,C2-AFM,A2-AFM and FM  configurations.}
	\label{fig:pd-j1-f}
\end{figure}

Our results are summarized in Fig.\ref{fig:pd-j1-f} which shows the case $J_1>0$. For large values of the ratios $J_2/J_1$ and $J_3/J_1$ we find the same AFM ground states as in the Ising case~\cite{crespo2013competing}. For smaller values of the ratios we find coplanar helical phases, that are, in general, incommensurate with the lattice. 
For $J_1<0$ the helical phases are absent and there are only the simpler AFM ground states, this implies that frustration is needed for the 
appearance of the helical phases. The large ratio ground states are relatively easy to 
identify and we focus on the helical states below.

We used two different approaches to find the ground states and compute their energy. First, we performed the well-known interaction matrix analysis~\cite{luttinger1946ltz,litvin1974ltz} that is briefly presented in Appendix\ref{sec:int-mat}. Second, we used numerical minimization which employs local field quench method.~\cite{walker1980computer} This method consists in starting from some, possibly random, initial configuration and minimizing the energy of the system by iteratively aligning every spin with its local field. This simple algorithm has proved its efficiency in the case of clean (without disorder) systems even in presence of frustration.~\cite{huber1993antiferromagnetic,henley2001effective} We have found that the lowest eigenvalue of the interaction matrix and the lowest energy reached by numerical minimization agree well with each other for all the couplings values considered. We concluded therefore that the lowest eigenvalue of the interaction matrix corresponds to the true ground state as it gives the 
lowest possible energy per site.~\cite{litvin1974ltz}  However because of the non-Bravais nature of the Hollandite lattice, it is not straightforward to construct the ground state spin configuration from the   wave vector $\vec{ q}$ which minimizes the interaction matrix. The reason is that in a lattice with a basis, usually it is not possible to make any three-component linear combination of  the lowest eigenmodes of interaction matrix (for $\vec{q}$) preserving the unit-length constraint for the spins. To construct the ground states, we have examined an equivalent auxiliary interaction matrix (see the Appendix\ref{sec:int-mat}). This method allows to find the exact expressions for the energy per site in terms of the couplings $J_k$'s. By using this procedure we obtain a one dimensional wave vector $\vec{q}=q$ which gives the ground state energy that coincides with the  lowest eigenvalue of the interaction matrix 
obtained from Eq.~\eqref{eqn:jq} and the lowest energy obtained in the numerical simulation. By associating an angle $\phi$ to the wave vector $q$, we find that the ground state energy can be written as (see Appendix\ref{sec:int-mat}):
\begin{gather}
	\label{eqn:siten}
	E_{GS} = J_1 \cos 2\phi + (|J_2|+ 2 |J_3|) \cos \phi.
\end{gather}

\begin{figure}[!t]
\includegraphics[width=0.9\columnwidth]{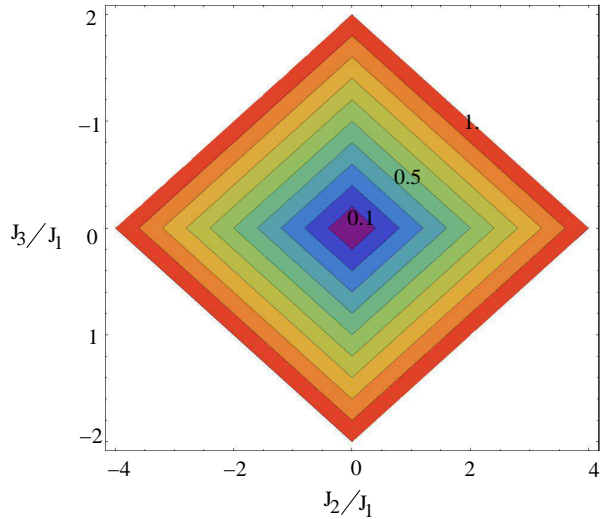}
	\caption{(Color online). Contour plot of the value $\phi$ as a function of the couplings. Each contour line refers to a fraction of 
	$\pi$, three values are represented $0.1\pi$, $0.5\pi$ and 1.$\pi$. The ground state energy with same values of $\phi$ is identical in each quadrant, however  the ground state spin configuration in each quadrant for a given $\phi$ is different as  explained in the text.}
	\label{fig:phi}
\end{figure}

Here $2\phi$ is the angle between spins connected by a $J_1$ bond in the $y$-direction and the angle $\phi$ is related to the angle between
neighboring spins joined by $J_2$ and $J_3$ bonds. The precise relation is specified below when we classify the possible helical states. 
As a consequence of Eq.~\eqref{eqn:siten}, we find a non-commensurate helical spin configuration in the $y$ direction with a rotation angle
of $2\phi$ as shown in Fig.\ref{fig:helical}. Therefore, the angle $\phi$ measures the helicity of the spin configuration.
The angle $\phi$ that minimizes the energy~\eqref{eqn:siten} is given by
\begin{gather}
	\label{eqn:phis}
	\cos\phi = - \frac{|J_2| + 2|J_3|}{4J_1}.
\end{gather}
Substituting the above value of $\cos\phi$, we find the following final expression for the ground state energy in the helical spin configuration,
\begin{gather}
	\label{eqn:fings}
	E_\text{GS} = -J_1 - \frac{(|J_2| + 2|J_3|)^2}{8J_1}.
\end{gather}

The condition for the angle $-1 < \cos \phi < 1$ implies that the above expression for site energy is valid only when Eqn.~\eqref{eqn:phis} has a solution for $\phi$, i.e $\frac{|J_2|}{4J_1} + 2\frac{|J_3|}{4J_1} \leq 1$. This condition determines the phase boundary
between helical spin states and the simple AFM ground states (see Fig.\ref{fig:pd-j1-f}). 
As both the energy and the angle $\phi$ depend on $J_2$ and $J_3$ only through the quantity $|J_2| + 2|J_3|$, we have that for each value of
this quantity there is a segment in the phase diagram where the energy and the spin configurations are exactly the same for all the points 
in a given quadrant.   This can be seen in Fig.\ref{fig:phi} where we show a contour plot of the value $\phi$ as a function of the couplings.
As can be seen from this plot the value of $\phi$ increases continuously from $0$ to $\pi$ forming a rhombus in phase space for a given
value of $\phi$. In general the spin configurations for a given value of $\phi$ will correspond to a non-commensurate co-planar spin 
arrangement (only for some specific values of $J_2$ and $J_3$, the angle $\phi$ is commensurate). It is worth to notice that we get an 
incommensurate ground state in the $J$-isotropic case where ($J_1 = J_2 = J_3$), for which we have $\phi \approx 2.42$ radians, very different from
the triangular lattice where there is an ordered ground state with  $2\pi/3$
 angle between them.  It is also important to note that the plane of polarization of the spins is arbitrary, 
as the Hamiltonian is invariant under global rotations of the spins.

The detailed analysis (see Appendix\ref{sec:int-mat} for more details) reveals that there are four different helical spin arrangements that appear for four different combinations of the signs of $J_2$ and $J_3$. We also notice that, as the angle $\phi$ increases continuously from $0$ to $\pi$, each of these states eventually transforms into one and only one of the AFM states found in the large $J_2/J_1$, $J_3/J_1$  limit. Therefore, we use similar notations for the four helical ground states as for the corresponding  AFM/FM state to which it is connected: C-H, C2-H, A2-H  and F-H phase,  the letter `H' refers to the helical nature of the ground states. In order to describe the spin arrangement in each of these helical ground states,  it is useful to label a given site with index $n$ in a specific $J_1$-chain (namely the $y$-axis as defined in the caption of Fig.\ref{fig:MnO2-interactions}). As we have mentioned above, the spins rotate helically in the lattice $y$-direction with a rotation angle of $2\phi$ (see Fig.\ref{fig:helical}). 
However, spins on different $J_1$-chains, connected by $J_2$ and $J_3$ bonds, can have different spin arrangements. To illustrate this difference and to describe the spin configurations, it is useful to divide all the $J_1$ into $\alpha$ chains, that are the chains with an odd sublattice number as given in Fig.\ref{fig:MnO2-lattice}(a), and $\beta$ chains for even sublattices. 

\begin{figure}
	\includegraphics[width=0.8\columnwidth]{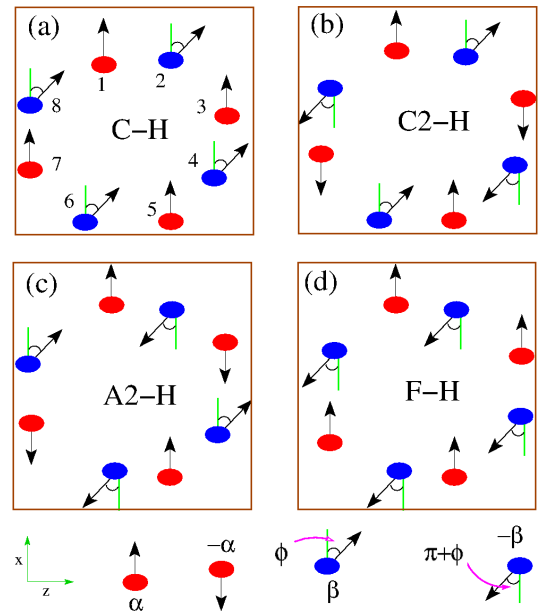}
	\caption{(Color online) Spin arrangement on the $x-z$ plane of each incommensurate helical phase. (a) C-H where the angle between the spins on nearest $J_1$-chains named $\alpha$,$\beta$ is $\phi$ (b) C2-H where the $J_1$-chains inside the unit cell are arranged $\alpha$,$\beta$,$-\alpha$,$-\beta$ and the angle between them are $\phi$ if the sign does not change and $\pi+\phi$ if the sing change (c) A2-H here the  $J_1$-chains are arranged $\alpha$,$-\beta$,$-\alpha$,$\beta$ and also the angle between them are $\phi$ if the sign does not change and $\pi+\phi$ if the sign change (d) F-H here we have $\alpha$,$-\beta$ with an angle of $\pi+\phi$.}
	\label{fig:xzhelical}
\end{figure}

We now describe four different ground states. In all these helical phases, the ground states correspond to co-planar spin arrangements with the spins in the $\etx-\etz$ plane and the pitch of the chirality advances towards the $y$-axis. The choice of the polarization plane is arbitrary. We start with the C-H phase ($J_2 > 0$ and $J_3>0$): The spins on the $\alpha$ and $\beta$ chains are parametrized as follows
\begin{gather}
	\label{eqn:alpha0}
	\vs_n^{~\alpha}  = \cos(2n\phi )\etx + \sin(2n\phi )\etz\\
	\label{eqn:beta0}
	\vs_n^{~\beta} = \cos((2n + 1)\phi)\etx + \sin((2n + 1)\phi )\etz.
\end{gather}
Here $\etx$ and $\etz$ are unit vectors in the $x$ and $z$ spatial directions, $\vs_n^{~\alpha}$ and $\vs_n^{~\beta}$ are spins on sites $n$ in the $\alpha$ and $\beta$ chains respectively with $n=0,1,2,3,...$. Then the arrangement of the spins in the C-H phase will be $\vs_n^{~\alpha}$ for odd sublattice number and $\vs_n^{~\beta}$ for even sublattice number as shown in Fig.\ref{fig:xzhelical}(a). It is straightforward to check that $\vs_n^{~\alpha}\cdot \vs_n^{~\beta} = \cos\phi$ and the angle between two spins in different $J_1$-chains is always $\phi$. The ground states for the other choices of the signs of $J_2$ and $J_3$ are constructed from the C-H ground state as we now show. For the C2-H family of ground states, we have $J_2 < 0$ (negated with respect to C-H case) and $J_3 > 0$. We obtain the ground state energy if  we set $\vs_{n(J_2)}^{~\alpha}\cdot\vs_{n(J_2)}^{~\beta} = -\cos\phi = \cos(\pi + \phi)$ and $\vs_{n(J_3)}^{~\alpha}\cdot\vs_{n(J_3)}^{~\beta} = \cos\phi$ where $\vs_{n(J_2)}^{{~\alpha}(
{\beta})}$ and $\vs_{n(J_3)}^{{~\alpha}({\beta})}$ are the two nearest-neighbor spins of the site $n$ of the $\alpha$ and $\beta$ chains connected to it by the $J_2$ or $J_3$ bonds respectively. Therefore, if two $J_1$-chains are linked by $J_2$ bonds the angle between the spins in the chains is $\pi + \phi$, while spins of the chains linked by $J_3$ bonds have an angle of $\phi$. Therefore, there are $4$ different $J_1$-chains per unit cell: $\alpha$,$\beta$,$-\alpha$,$-\beta$ as shown in Fig.\ref{fig:xzhelical}(b). The minus sign denotes a spin flip. Similar reasoning applies to the A2-H phase where $J_2 > 0$ and $J_3 < 0$: the angle between two spins linked by $J_2$ bond is $\phi$ and spins linked by $J_3$ bond  form an angle $\pi + \phi$. The arrangement of the $J_1$-chains in the A2-H ground state is $\alpha$,$-\beta$,$-\alpha$,$\beta$ as shown in Fig.\ref{fig:xzhelical}(c). Finally in the F-H ground state where both $J_2 < 0$ and $J_3 < 0$, the angle between two spins in different $J_1$-chains is 
always $\pi + \phi$ and the order of the $J_1$-chains is $\alpha$,$-\beta$ as shown in Fig.\ref{fig:xzhelical}(d). It is worth to mention that the definition given above for the $\alpha$ and $\beta$ index to the odd and even sublattices is not unique. In fact the definition of $\alpha$ and $\beta$ chains could be interchanged by adding a constant angle $-\theta$ in the argument of the $\sin$ and $\cos$ terms appearing in above Eqs.~\eqref{eqn:alpha0}-\eqref{eqn:beta0}. However this fact does not correspond to any macroscopic degeneracy of the ground state spin configuration.

It is of interest to compare the distribution of angles between spins within a given triangle for the Hollandite lattice to that of
other 2d triangular lattices or ladders.  For a given spin in a triangle in the Hollandite lattice, if $\phi^{\prime}$ and $\psi$
denote the angle of the spins joined by $J_1$ bonds and $J_2$ (or $J_3$) respectively, then we obtain the condition 
$\phi^{\prime}=2 \phi$ for any values of $J_k$'s and $\psi=\phi$ or $\psi=\pi + \phi$ depending on the sign of $J_2$ and $J_3$. The relation 
$\phi^{\prime}=2 \phi$ has also been observed in other triangular lattices where the interaction strengths along different directions are
taken different.~\cite{weihong1999trafm}. It is also instructive to compare the resulting phase diagram with that of Ising spins on the same lattice.~\cite{crespo2013competing} For $J_1<0$ (ferromagnetic coupling) the interactions are not frustrated, and the two models have exactly the same ground states, as expected. The $J_1>0$ case is frustrated: If $|J_2|/J_1,|J_3|/J_1\gg1$ the ground states are again the same: C-, C2- and A2-AFM. For small values of $|J_2|/J_1,|J_3|/J_1$ ($\frac{|J_2|}{4J_1} + 2\frac{|J_3|}{4J_1} \leq 1$) differences appear between the Heisenberg and the Ising models: the Heisenberg model develops different incommensurate helical spin ground states, the Ising model still has the same collinear ground states. When one goes closer to the origin of phase space, $\frac{|J_2|}{2J_1} + \frac{|J_3|}{J_1}\leq 1$, the Ising model enters the correlated, geometrically frustrated phase, see Fig.~4 in Ref.~\onlinecite{crespo2013competing}). This phase is absent in the Heisenberg model, in which there are four different incommensurate helical 
ground states. The only degeneracy of these ground states is due to global spin rotations. We remark that the area of the correlated frustrated phase of the Ising model is smaller than that of the helical phases, implying that the Ising model underestimates the degree of geometrical frustration present in real materials, that are better described by the Heisenberg model. For the Ising model, the site energy $E_{GS}$
of the correlated, geometrically frustrated phase, does not depend on $J_2$ and $J_3$, unlike the present case, as seen from Eq. \ref{eqn:fings}. Also we note that there exists no spin configuration in the highly degenerate ground state manifold of the correlated, geometrically frustrated phase of Ising model which can continuously evolve to the respective collinear phase in the large $|J_2|/J_1$ and $|J_3|/J_1$ limit. However for the present model the helical spin configuration in each quadrant  of the phase diagram is  continuously connected to the respective collinear phase in the large $|J_2|/J_1$ and $|J_3|/J_1$ limit.

\section{Magnetic structure factor}
\label{sec:swt-et}

Experimental results suggest that helical magnetic ordering might be present at low temperatures in some Hollandite
compounds~\cite{sato1999charge,sato1997tunnel} but no details regarding the spin configurations is mentioned. In this section we discuss possible experimental signatures of the helical ground states presented in the previous section. The usual way to determine the spin configuration in a magnetic material is to perform neutron diffraction experiments.~\cite{lovesey1984neutron} In these experiments, the main quantity is the magnetic structure factor $\vec{F}_M(\vec{Q})$, defined as the Fourier transform of the magnetic structure:
\begin{gather}
	\label{eqn:magform}
	F_M(\vec{Q}) = \frac{1}{\sqrt{N_t}}\sum_{l=1}^{N_t} e^{\vec{Q}. \vec{r}_l}\vec{S}(\vec{r}_l).
\end{gather}
Here $\vec{Q}$ refers to the scattering momenta and $N_t$ is the total number of sites in the lattice. We note that $F_M(\vec{Q})$ is in
general a complex vector quantity. The  magnetic form factor $F_M(\vec{Q})$, could be  re-written for the four different helical phases in 
the following compact way,

\begin{gather}
	\label{eqn:unit-str}
	F_M(\vec{Q}) = \sum_{i=1}^{N_{uc}} \big( \sum^{5,7}_{j \in 1,3} \vec{s}_{i,j} e^{i \vec{Q}. \vec{d}_j} + \sum^{6,8}_{i \in 2,4}  \vec{s}_{i,j} e^{i \vec{Q}. \vec{d}_j} \big) e^{\vec{Q}.\vec{R}_i}~~~~ 
\end{gather}
where $N_{uc}$ is the total number of unit cells. While writing Eq.~\eqref{eqn:unit-str}, we have used $\vec{r}_l= \vec{R}_i + \vec{d}_j$, where $\vec{R}_i$
denotes the position of the unit cell center and $\vec{d}_j$ denotes the sublattices within the unit cell. $\vec{s}_{ij}$ is the spin for the `$j$' sublattice in `$i$'th unit cell. The quantity within the parenthesis of Eq.~ \eqref{eqn:unit-str} is related to the so called magnetic unit cell form factor defined in a given x-z plane. 
\begin{gather}
	\mathcal{F}_{M}(\vec {Q})= \left(\sum^{5,7}_{j\in1,3} \vec{s}_{j} e^{i \vec{Q}. \vec{d}_j}+\sum^{6,8}_{i\in 2,4}  \vec{s}_{j} e^{i \vec{Q}. \vec{d}_j}\right)~~~~~
\end{gather}
In the above $\vec{s}_j$ denotes the relative spin orientations in a given unit cell as indicated in Fig.\ref{fig:xzhelical}. The $\vec{d}_j$ denotes the position of the sublattice $j$ within a given unit cell as explained in Appendix\ref{sec:int-mat}.

\begin{figure*}
	\includegraphics[height=45mm,width=180mm]{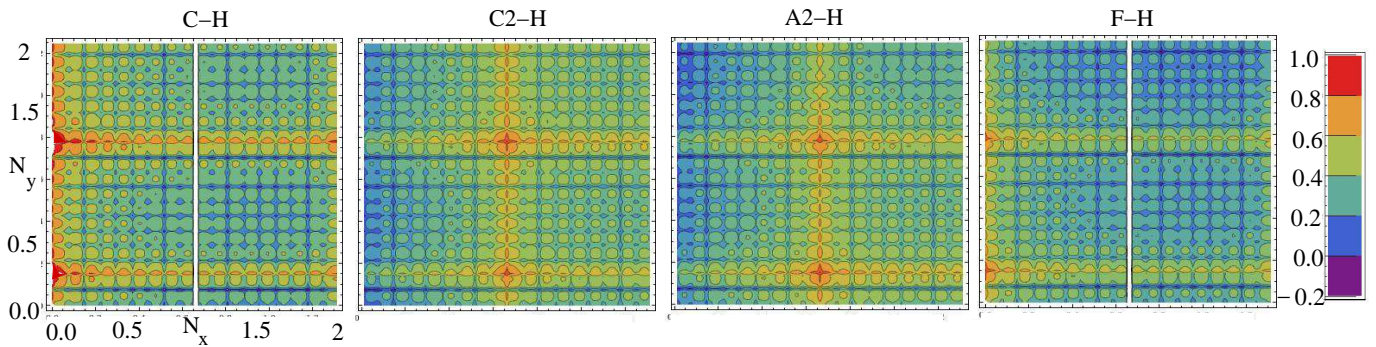}
	\caption{(Color online) A contour plot of  ${\rm{Log}}_{10}(|F_M(\vec{Q}_{max})|/|F_M(\vec{Q}_M)|$ (see text for details) has been given in the plane of $Q_y$ and $Q_{x}$. $Q_y$ and $Q_x$ are given by $\frac{2 \pi}{a} N_x$ and $\frac{2 \pi}{c} N_y$ respectively and the length of  each axis has been taken as twice the reciprocal unit vector along respective axis. The axes are labeled by the values of $N_x$ and $N_y$  which takes values `$0$' to `2'. `$a$' and `$c$' denotes the reciprocal lattice vectors along the $x$ and $y$  directions respectively. The bright regions indicate peaks i.e  higher values of $|F_M(\vec{Q}_M)|$. For the parameters $J_1=1.0,~|J_2|=2.0$ and $|J_3|=0.45$ the $\phi$ is 2.38 radians. The bright  lines parallel to $Q_x$ axis appear for $Q_y \sim 0.13$ and  $Q_y \sim 0.79$. These values of $Q_y$ corresponds to   $Q_y=2 \phi - \frac{14 \pi}{c}$ and $Q_y=2 \phi - \frac{12 \pi}{c}$ respectively.}
	\label{fig:mag-form}
\end{figure*}

To illustrate the qualitative difference present in  the four different helical spin configurations we choose  $J_1= 1.0,~J_2= \pm 2.0$ and
$J_3=\pm 0.45$. The choice of the magnitude of the $J_2$ and $J_3$ is for illustrative purpose only and the arguments presented here could 
easily be extended for the cases when the complete specification of the $J_i$'s are available. There are two key quantities related to the 
ground states of the helical phases. First, the angle $\phi$ that specifies the rotation angle of spins  along the $y$-axis. The second quantity
is the relative orientation of the sublattice spins in a given unit cell as shown in Fig.\ref{fig:xzhelical}. The information regarding the
angle $\phi$ is encoded in the $y$-component of the scattering vector $\vec{Q}$, and the relative orientation of spins in different helical 
phases can be read off from  $x$- and $z$-components of $\vec{Q}$. The angle $\phi$ could easily be inferred by detecting the maximum of 
$|F_M(\vec{Q})|$ as a function of $\vec{Q}_y$. To illustrate this, we have presented a $2$-dimensional contour plot of $|F_M(\vec{Q})|$ in
$x-y$ plane as shown in Fig.\ref{fig:mag-form}. For the parameters, $J_1=1.0,~|J_2|=2.0$ and $|J_3|=0.45$, angle $\phi$ (following 
Eq.~\eqref{eqn:phis})  is equal to $2.38$ radians. A straightforward evaluation of Eq.~\eqref{eqn:unit-str}, yields that one expects
peaks to appear at $Q_y = \pm \big( \frac{2 \pi N}{c} - 2 \phi \big)$, but the peak heights at a given $Q_x$ is not, in general, the same for 
each phase. These constitute the key features of the magnetic form factor and could be used to distinguish the four helical phases.

We note that the position of the maximum of $|F_M(\vec{Q})|$, i.e, $|F_M(\vec{Q})|_{max}$ is different in each phase. 
The corresponding $\vec{Q}$ is named $\vec{Q}_{max}$ and  $|F_M(\vec{Q})|_{max}$ is simply named $|F_M(\vec{Q}_{max})|$ here in after. 
The position of $\vec{Q}_{max}$ and the relative height of other peaks compared to $|F_M(\vec{Q}_{max})|$ in a given phase is 
characteristics of that phase and could be used to distinguish each phase from another. This is particularly important as in a given experiment 
the obtained value of $|F_M(\vec{Q}_{max})|$ is completely arbitrary and can not be distinguished for each phase but the relative height of 
other peaks compared to $|F_M(\vec{Q}_{max})|$ could be determined. For this reason we have plotted in Fig.\ref{fig:mag-form} 
${\rm{Log}}_{10}$ $(10 |F_M(\vec{Q}_{max})|/|F_M(\vec{Q})|)$ in $Q_x$-$Q_y$ plane. We notice from Fig.\ref{fig:mag-form} that for 
C-H and F-H phases the  two largest peaks appear at the  wave vectors $\vec{Q}=(0,- \big( \frac{12 \pi}{c} - 2 \phi \big),0)$ and
$\vec{Q}=(0,- \big( \frac{14 \pi }{c} - 2 \phi \big),0)$ respectively. However if we compare the ratio of the heights of these two peaks we find that they take different 
values in each phase. This is also true for the C2-H and A2-H phases where the two largest peaks appear at 
$\vec{Q}=(\frac{2 \pi}{a},- \big( \frac{12 \pi }{c} - 2 \phi \big), 0)$ and  $\vec{Q}=(\frac{2 \pi}{a},-\big( \frac{14 \pi }{c} - 2 \phi \big), 0)$ respectively.
In Table~\ref{table:nonlin}, we have represented the ratio of the relative peak heights at different pairs of points, $(N_1,(2 \phi-N_2))$ 
(where $\vec{Q}= (\frac{2 \pi}{a}N_1, -\big( \frac{2 N_2 \pi }{c} - 2 \phi \big),0)$) to demonstrate the qualitative difference between the four helical phases. In
these points $|F_M(\vec{Q})|$ is directly proportional to $|\mathcal{F}(\vec{Q})|$ and thus one can compare the theoretically obtained
values for $\rm{Log}_{10}$ $(10 |F_M(\vec{Q}_{max})|/|F_M(\vec{Q})|)$ to that of experimentally obtained values to arrive at the correct phases.
We note that for the C-H phase and the F-H phase there is a white line at $Q_x=\frac{2 \pi}{a}$. This is due to the fact that for $\vec{Q}=({\frac{2 \pi}{a}, Q_y,0})$, the magnetic unit cell structure factor  $|\mathcal{F}_{M}(\vec{Q})|$ is identically zero for any values of $Q_y$ and $\phi$ (the contribution from the odd and even sublattices are zero separately). The $\rm{Log}_{10}$ plot used in the Fig.\ref{fig:mag-form} makes  $\rm{Log}_{10}
$ $(10 |F_M(\vec{Q}_{max})|/|F_M(\vec{Q})|)$ a large negative number which is outside the  numerical range shown by the color code 
in Fig.\ref{fig:mag-form} and hence represented by white line instead. This particular sublattice symmetry is broken for the
C2-H and A2-H phases by  different distribution of sublattice spins as  evident from Fig.\ref{fig:xzhelical}.
 
\begin{table}[ht]
	\caption{Comparative ratios of $|F_M(\vec{Q})|$ (or $|\mathcal{F}_{M}(\vec{Q})|$) at symmetric points which are represented by a pair of numbers. The first column represent different  helical phases and the other columns represent $|F_M(\vec{Q})|$ or ($|\mathcal{F}_{M}({\vec{Q}})|$) scaled by maximum of $|F_M(\vec{Q})|$ (or $|\mathcal{F}_{M}({\vec{Q}})|$).} 
\centering 
\begin{tabular}{c c c c c c c} 
\hline\hline 
Phase & (0,2$\phi$-7) & (1,2$\phi$-7) & (2,2$\phi$-7) & (0,2$\phi$-6)& (1,2$\phi$-6)& (2,2$\phi$-6) \\ [0.5ex] 
\hline 
C-H & 0.5618 & 0 & 0.0597 &1&0&0.1063\\ 
C2-H & 0 & 0.5618 & 0 &0&1&0\\
A2-H & 0 & 1 & 0 &0&0.5933&0\\
F-H & 1 & 0 & 0.1063 &0.6566&0&0.0698\\[1ex]
\hline 
\end{tabular}
	\label{table:nonlin} 
\end{table}

One can perform a similar study of $F_M(\vec{Q})$ in $Q_x$-$Q_z$ plane also to differentiate the four different helical phases, however we 
refrain from discussing that as it will add little to what has already been explained here. Having discussed in detail the experimental
realizations of the ground state spin configurations, we discuss the  ground state  magnetization and zero field susceptibility. The
ground state magnetization ($m_{\mu}$= $\sum_i s_{i,\mu})$ is zero as can be found directly from Eqs.~\eqref{eqn:alpha0}-\eqref{eqn:beta0}.
The $T=0$, zero magnetic field susceptibility tensor is given by,
\begin{eqnarray}
\chi_{\mu,\lambda} &=&\frac{1}{N}\big( \sum_{i,j} \langle s_{i,\mu} s_{j,\lambda} \rangle - \langle s_{i,\mu} \rangle  \langle s_{j,\lambda} \rangle\big) \nonumber \\
	&=& \chi ~ \delta_{\mu,\lambda},~~ \mu, \lambda \in x,z
\end{eqnarray}
In the above $\chi=0.5$  and $\delta_{\mu,\lambda}$ is the Kronecker delta, as we have taken the plane of spin 
configurations as the $x-z$ plane. The vanishing ground state magnetization and finite zero temperature susceptibility are consistent
with experimental values observed in a previous study.~\cite{sato1997tunnel}

\section{Effect of magnetic field}
\label{magfield-effect}

In this section we consider the effect of a uniform magnetic field. As mentioned in Sec.\ref{sec:clean}, the Hamiltonian is symmetric under 
global spin rotations. As a consequence, the plane of polarization for the helical spin configurations is arbitrary and the ground state 
energy is independent of the choice of the plane of polarization. For concreteness we assume the same $x$-$z$ plane we used in the previous sections. 
Therefore a parallel field ($h_\parallel$) means parallel to the director of the helix, i.e. $y$-direction in our case.
For small parallel uniform fields the Hamiltonian reads:
\begin{gather}
	\label{eqn:ymag}
	H = H_0 + h_\parallel\sum_{i=1}^{N_t} s_{y,i},
\end{gather}
where $H_0$ refers to the Hamiltonian given in Eq.~\eqref{eqn:H} and $s_{y,i}$ represents the $y$-components of $\vs_i$ at site $i$. We assume
the helical ground state is only perturbed slightly away from the zero field state, so that we can decompose every spin
$\vs_i=\sqrt(1-\Delta^2)\vs_{0,i} - \Delta\ety$ into zero field value $\vs_0$ and a small perturbation $\ety$, the value of $\Delta$ that minimizes
Eq.\ref{eqn:ymag} is $\Delta= \frac{h_\parallel}{4(J_1+J_2+ 2J_3 - E_{GS} )}$. Then the ground state energy in the presence of a perpendicular magnetic field equals to
\begin{gather}
	\label{eqn:magen}
	E = E_{GS}- \frac{h^2_\parallel}{4(J_1+J_2+ 2J_3 - E_{GS} )}.
\end{gather} 
As expected we find that the classical ground state energy is minimized in the presence of a magnetic field.  If we take the $J$-isotropic limit $J_1=J_2=J_3$, then from Eq.~\eqref{eqn:magen} we find the ground state energy for classical AFM 2D triangular lattice model in the presence of a magnetic field.~\cite{hikaru1985magfield} The susceptibility $\chi_\parallel$ is readily computed and is given by
\begin{gather}
	\label{eqn:chi}
	\chi_\parallel = \frac{1}{2(J_1+J_2+ 2J_3 - E_{GS} )} > 0.
\end{gather}
Therefore for a finite parallel magnetic field, we obtain $\chi_{\parallel} > 0$ at zero temperature in agreement with experimental results.~\cite{sato1999charge,sato1997tunnel} The above equation~\eqref{eqn:chi} is valid for all values of the parallel field such that $h_\parallel\leq h_\parallel^c$, where $h_\parallel^c$ is the critical field beyond which the helical order breaks down and the system becomes ferromagnetically ordered along the applied field. The critical field $h_\parallel^c$ can be calculated as 
\begin{gather}
	\label{eqn:crh}
	h_y^c = 2 (J_1+ J_2 + 2 J_3 -E_{GS}).
\end{gather}
Note that though the ground state state energy given by Eq.~\eqref{eqn:fings}, is independent of the sign of $J_i$, the susceptibility $\chi_\parallel$ and the value of the critical magnetic field $h_\parallel^c$ do depend on the sign of $J_k$.

\section{Conclusions}
\label{sec:conclusions}

We have studied a classical Heisenberg model with $J$-anisotropic couplings on the Hollandite lattice. The lattice is a good approximation for the structure of 
certain transition metal oxides, such as $\alpha$-MnO$_2$. By using both the interaction matrix method and the local field quench numerical method
we construct the phase diagram at zero temperature as a function of the couplings. We found a rich phase diagram with $8$ different phases,
depending on the signs and mutual strengths of the couplings $J_1,J_2,J_3$. The sign of $J_1$ has a strong effect on the phase diagram: 
for $J_1>0$ only collinear antiferromagnetic phases are present, while for $J_1<0$ these antiferromagnetic (AFM) phases are only stable for
large values of the ratios of $J_2/J_1$ and $J_3/J_1$. For smaller values of the ratios we find helical phases, with a pitch that in general
is not commensurate with the lattice. It is incommensurate even in the $J$-isotropic limit $J_1=J_2=J_3$, in contrast to the known $120^{\circ}$ spin distribution of 2D and 3D triangular lattices. The fact that the helical 
phases appear for  any $J_1>0$ indicates that frustration is the crucial ingredient for the existence of the helical order in this system. 
We showed that the four helical phases are connected to only one of the collinear antiferromagnetic phases that appear at large values of the ratios $J_2/J_1$ and $J_3/J_1$. 

By comparing results presented here with the previous results obtained for Ising spins we show that though for  large ratios of $J_2/J_1$ and $J_3/J_1$ the Ising solution is recovered, as expected, for small values the Ising model fails to
measure the frustration present in the lattice. We have also analyzed in detail the magnetic structure factor of each of these helical phases. Our analysis explains  the essential differences that magnetic structure factor possesses in each helical phase. We expect that the general argument outlined regarding the structure of magnetic structure factor will be useful to distinguish each phase in neutron diffraction  experiments. These results are of importance for compounds like K$_{1.5}$(H$_3$O)$_x$Mn$_8$O$_{16}$ and K$_{0.15}$MnO$_2$ where the helical 
arrangement have been suggested previously ~\cite{sato1999charge,sato1997tunnel} at low temperature. Finally we have studied the effect of an
external magnetic field perpendicular to the plane of polarization of the helical spin configurations and calculated the magnetic susceptibility for such external magnetic field. The critical external magnetic field  for which the helical order breaks down is also provided.

\begin{acknowledgments}
	We thank Mikhail Feigelman and Roderick M\"ossner for useful discussions.
\end{acknowledgments}

\appendix

\section{Interaction matrix method}
\label{sec:int-mat}

In this appendix we present briefly the standard interaction matrix analysis and the extended interaction matrix method, that we developed to construct the helical ground states.

\begin{figure}[!t]
	\includegraphics[width=0.99\columnwidth]{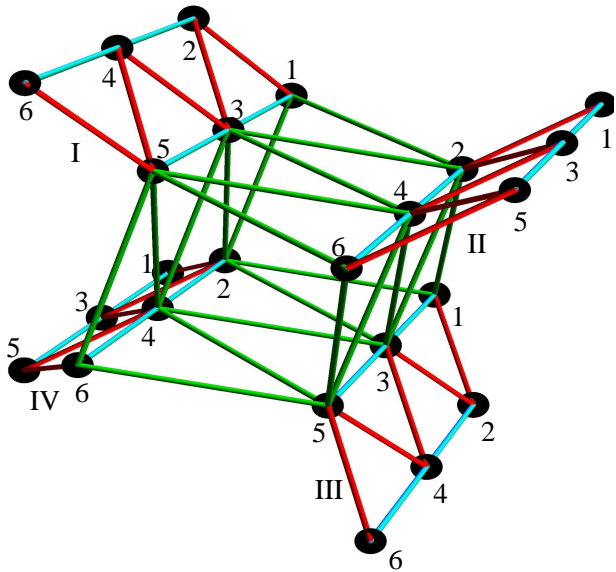}
	\caption{(Color online) A part of the Hollandite lattice where four $J_1-J_2$ ladders are connected to form a $4$ site channel. The letters $I$ to $IV$ denote the four different $J_1-J_2$ ladders extending along the $y$-directions. The numbers $1,2,3,....$ represent the indexing of sites in each ladder.}
	\label{fig:hollandite-ladder}
\end{figure}

To calculate the interaction matrix, we start by enumerating the sites of unit cell as shown in Fig.\ref{fig:MnO2-lattice} (a). The coordinates of a given site in Hollandite lattice are given by a vector $\vec{R}_{i,m}$ where $i$ denotes the unit cell and $m$ refers to the sublattice. The $\vec{R}_{i,m}$ is written as $\vec{r}_i + \vec{d}_m$ where $\vec{r}_i$ is the position of the center of the unit cell and $\vec{d}_m$ the relative position of each unit cell atom with respect to the center, with $m=1,2,\ldots,8$. Within this $8$-atom unit cell, we can express the vectors $\vec{d}_m$ as functions of the lattice spacing. Defining $a_1$,$a_2$ and $a_3$ as distances between the neighbors connected by bonds $J_1,J_2$ and $J_3$ respectively, we find: 
\begin{eqnarray}
	&& \vec{d}_1 = -l_2 \vec{e}_x+ \frac{a_1}{2} \vec{e}_y+l_1 \vec{e}_z,~\vec{d}_2=l_2 \vec{e}_x+l_1 \vec{e}_z, \notag\\
	\label{eqn:subvec}
	&& \vec{d}_3 = l_1 \vec{e}_x+\frac{a_1}{2} \vec{e}_y +l_2 \vec{e}_{z},~\vec{d}_4=l_1 \vec{e}_x-l_2 \vec{e}_z,\\
	&& \vec{d}_5 = l_2 \vec{e}_x+ \frac{a_1}{2} \vec{e}_y-l_1 \vec{e}_z,~\vec{d}_6=-l_2 \vec{e}_x-l_1 \vec{e}_z, \notag\\
	&& \vec{d}_7 = -l_1 \vec{e}_x + \frac{a_1}{2} \vec{e}_y+l_2 \vec{e}_z,~\vec{d}_8=-l_1 \vec{e}_x+l_2 \vec{e}_z, \notag	
\end{eqnarray}
where the factors $l_i$'s are given by
\begin{eqnarray}
	&& l_1=l_2 + \frac{1}{\sqrt{2}}\sqrt{a_3^2- \frac{a_1^2}{4}}, ~~l_2= \frac{1}{2}\sqrt{a_2^2- \frac{a_1^2}{4}},
\end{eqnarray}
and $\vec{e}_\eta$ denotes the unit vector along the $\eta$ axis with $\eta=x,y,z$. It more convenient, however, to describe the Hollandite lattice using a unit cell containing just $4$ sites as shown in Fig.\ref{fig:MnO2-interactions}. The $4$ lattice points selected as a basis are drawn in blue in Fig.\ref{fig:MnO2-interactions}. The  translation vectors ($\vec{A}_1,\vec{A}_2,\vec{A}_3$) as functions of the bond distances are given by the following expressions:
\begin{eqnarray}
	\vec{A}_1 &=& a_1 \vec{e}_y\notag\\
	\vec{A}_2 &=& 2( l_1 + l_2) \vec{e}_x \\
	\vec{A}_3 &=& l_2 \vec{e}_x- \frac{a_1}{2} \vec{e}_y-l_1 \vec{e}_z.\notag
\end{eqnarray}
These vectors are shown in Fig.\ref{fig:MnO2-interactions} as pink arrows starting form the lattice point number $2$.  

Now we are in position to present the interaction matrix methods. First, we outline the usual interaction matrix method for classical spin
system.~\cite{luttinger1946ltz} as given below,
\begin{gather}
	J_{mk}(\vec{q}) = \frac{1}{N}\sum_{i,j} J_{im,jk} e^{i\vec{q}\cdot(\vec{R}_{im} - \vec{R}_{jk})},
\end{gather}
where $i$ and $j$ denote the index of the unit cells of the lattice, and $m$ and $k$ denote the site position in the unit cell. As there are only $4$ atoms in the unit cell, the interaction matrix is a $4\times4$ matrix. One can easily check that the symmetry of the Hollandite lattice, reduces the number of independent matrix elements to $4$. Referring to the selected $4$-site unit cell of the Hollandite lattice as described in Fig.\ref{fig:MnO2-interactions}, the independent matrix elements are given below,
\begin{align}
	J_{11}(q) & = 2 J_1 \cos(\vec{q}\cdot\vec{u}_{11})\notag\\
	J_{12}(q) & = J_2 e^{i\vec{q}\cdot\vec{u}_{12}} + J_2 e^{i\vec{q}\cdot(\vec{u}_{12} - \vec{u}_{11})}\notag\\
	J_{13}(q) & = J_3 e^{i\vec{q}\cdot \vec{u}_{13}} + J_3 e^{i\vec{q}\cdot(\vec{u}_{13} - \vec{u}_{11})}\notag\\
	J_{14}(q) & = J_3 e^{i\vec{q}\cdot\vec{u}_{14}} + J_3 e^{i\vec{q}\cdot(\vec{u}_{14} - \vec{u}_{11})},\notag
\end{align}
where $\vec{u}_{11}=\vec{A}_1$, $\vec{u}_{12}=\vec{d}_8 - \vec{d}_1$, $\vec{u}_{13}=\vec{d}_8 - \vec{d}_7$ and $\vec{u}_{14}=\vec{u}_1-\vec{d}_3 + \vec{d}_7$ using Eqs.~\eqref{eqn:subvec}. We have the following expressions for the interaction matrix,
\begin{gather}
	\label{eqn:jq}
	J(\vec{q}) = \left(
	\begin{array}{cccc}
		J_{11}(\vec{q}) & J_{12}(\vec{q}) & J_{13}(\vec{q}) & J_{14}(\vec{q})\\
		J_{12}(\vec{q}) & J_{11}(\vec{q}) & \frac{J_3}{J_2} J_{12}(\vec{q}) & J_{13}^*(\vec{q}) \\
		J_{13}^*(\vec{q}) & \frac{J_3}{J_2} J_{12}^*(\vec{q}) & J_{11}(\vec{q}) & J_{12}(\vec{q})\\
		J_{14}^*(\vec{q}) & J_{13}(\vec{q}) & J_{12}^*(\vec{q}) & J_{11}(\vec{q})
	\end{array}
	\right),
\end{gather}
where the asterisk denotes complex conjugation. 

Because of the non-Bravais nature of the Hollandite lattice, it is difficult to find the ground state spin configuration, even 
in the case when the value of $\vec{q}$ that minimizes the interaction matrix, is known. In order to solve this problem we developed
the following alternative approach. We notice that the Hollandite lattice can be described as a   collection of $J_1-J_2$ ladders
interconnected by the $J_3$ links (see Fig.\ref{fig:hollandite-ladder}). We next construct the interaction matrix of a given $J_1-J_2$
ladder and also the inter-ladder interaction matrix for the $J_3$ coupling. In this representation the wave vector $\vec{q}$ became one 
dimensional ($\vec{q} = q$). We begin by examining the minimum wave vector $q$ that minimizes 
the new interaction matrix for the collection of such $J_1-J_2$ ladders. We found that for the four $J_1-J_2$ ladder as shown in 
the Fig.\ref{fig:hollandite-ladder}, the $q$ which minimizes the interaction matrix, yields the lowest ground state site energy as
obtained by numerics and usual interaction matrix in Eq.~\eqref{eqn:jq}. The interaction matrix 
for these four $J_1-J_2 $ ladder system reduces to:
\begin{gather}
	\label{eqn:aimjq}
	J^{\prime}(q) = 
	\left(
	\begin{array}{cccc}
		J_0(q) & J_1(q) & 0 & J_1(q)\\
		J_1(q) & J_0(q) & J_1(q) & 0\\
		0 & J_1(q) & J_0(q) & J_1(q)\\
		J_1(q) & 0 & J_1(q) & J_0(q)
	\end{array}
	\right),
\end{gather}
where the diagonal terms $J_0(q)=J_1 \cos 2q + J_2 \cos q$ corresponds to interactions inside one ladder and the off-diagonal terms $J_1(q)=J_3 \cos q$ to inter-ladder interactions, for simplicity we have neglected the vector representation of $q$ as it is one-dimensional. The four eigenvalues obtained for $J^{\prime}(q)$ are $\pm (J_1 \cos 2q + J_2 \cos q)$ and $J_1 \cos 2q \pm (J_2+ 2 J_3) \cos q$. The lower eigenvalue is $ J_1 \cos 2q + (J_2+ 2 J_3) \cos q$ corresponding to Eq.~\eqref{eqn:siten}.


\end{document}